
 \magnification 1200
\hsize=5.9truein
\null\vskip 1cm
\centerline{{\bf CHERN-SIMONS  QUANTUM FIELD
REPRESENTATION  } } \centerline{{\bf OF PLANAR
FERROMAGNETS } } \vskip 1cm
 \centerline{L. Martina, O. K. Pashaev $^{*)}$ and G. Soliani
}  \vskip 1cm \centerline{{\it Dipartimento di Fisica
dell'Universita' and Sezione INFN,  73100 Lecce, Italy} }
\centerline{$^{*)}${\it Joint Institute
for Nuclear Research,141980 Dubna, Russia} } \vskip 2cm
\par \noindent{\it  We  represent
 the two - dimensional planar classical continuous
Heisenberg spin model  as a constrained Chern-Simons
gauged nonlinear Schr\"odinger system. The hamiltonian
structure of the model is studied, allowing the
quantization of  the theory   by the gauge invariant
approach.  A preliminary study of the quantum states is
displayed and  several physical consequences in terms of
anyons are discussed.} \vskip 1cm
 \par \medskip \noindent  {\bf 1.} $\qquad$ Chern-Simons
(CS) gauged models have  recently attracted a great
attention  in  the description of  several phenomena in
condensed matter physics as
 the Fractional Quantum  Hall  Effect  (FQHE)  and
the High Temperature Superconductivity [1, 2].  In this
context, models of the  Landau-Ginzburg  type   have been
proposed as
  effective  field-theories
[3, 4].  In the
stationary limit,  soliton solutions
to  the self-dual reductions of these models appear.
These can be exploited  to represent
Laughlin's  quasiparticles  and  anyons  [5]. On the other
hand,  the  present authors
 showed   that  the  self-dual CS  system  can
be obtained,  in  the tangent space representation, from
the stationary two-dimensional Landau - Lifshitz equation
(LLE)  [6].  This is the classical  continuous
isotropic version of the  Heisenberg model,  obtained
by  the
 discrete quantum version  in the framework of the
coherent state representation [7]. In particular,  soliton
solutions  of the former system  turn out to be   related
to  magnetic vortices of the  LLE.  More generally,  the LLE
in   2+1  dimensions takes  the form of a constrained
nonlinear Schr\"odinger system (cNLS) for two  charged
matter  fields,  coupled to a   CS  gauge  potential [8].  This
correspondence suggests an explanation  to the
phenemenological analogy  between the   behaviour  of
magnetic vortices [9] and Hall particles in an external
magnetic field [10].  \par  Sec. 2 is devoted to a brief
review of the Landau - Lifshitz model  in 2+1 dimensions
and its
 tangent space formulation,  that is the  CS gauged cNLS
system. Furthermore,  the conservation of some associated
physical quantities is discussed. In Sec. 3 we study  the
lagrangian formulation and the hamiltonian structure of
the CS gauged cNLS system.    Finally, in Sec. 4  we
apply   the method  of gauge
invariant quantization to the above mentioned system.
\par\medskip \noindent  {\bf 2.}  $\qquad$  In 2+1
dimensions the LLE  for   the local
magnetization  ${\vec S}$ reads $$ \vec S_{t} = \vec S
\times  \nabla^{2} \vec S , \eqno(1)$$
  where $  \vec S $ belongs to the  2-dimensional sphere
${\cal S}^{2}$.
 In terms of the
 stereographic projection  $$ S_{+}= S_{1} + {\mit i}\; S_{2}
=  2\,{\zeta  \over  {1+\left|\zeta \;\right|^{\,2}}} ,  \qquad
S_{3}={{1-\left|\zeta\; \right|^{2}} \over {
1+\left|\zeta \;\right|^{2}}} \eqno(2)$$  Eq. (1) becomes
$${1 \over 4}\;i\;{\zeta }_{\;t}+{\zeta }_{\;z\;\overline{z}}=
2{{\zeta }_{\;z}\;{\zeta }_{\;\overline{z}} \over
1+{\left|{\zeta }\;\right|}^{2}}\overline{\zeta }\qquad (z =
x_{1} + i x_{2}, \; \overline{z} = x_{1} -  i x_{2}  ).
\eqno(3)$$
With   ferromagnetic boundary conditions, that is
\hbox{${\vec S}({\bf r},t) \rightarrow  (0,0,-1)$ for $|{\bf
r} |
\rightarrow \infty $},   the $\left( x_{1}, x_{2} \right)$
plane is compactified. Thus       ${\vec S}$ defines
maps of
 ${\cal S}^{2}\rightarrow  {\cal S}^{2}$,  which are
partitioned into homotopy classes,  labeled
by  the integer-valued
 topological  charge
 $$ Q = {1\over{4 \pi}} \int{\vec S} \cdot (\partial
_{1}{\vec S \times  \partial }_{2}{\vec S})\;d^{2}x  =  {1
\over \pi }\int_{}^{}{{\left|{{\zeta
}_{\;\overline{z}}}\;\right|}^{2}-{\left|{{\zeta
}_{\;z}}\;\right|}^{2} \over {\left({1+{\left|{\zeta
}\;\right|}^{2}}\right)}^{2}}\;d^{2}{x} .\eqno(4)$$   The
energy of the system takes the form  $$W={1 \over
2}\int_{}^{}{\left({\nabla \vec{S}}\;\right)}^{2}\;
d^{2}{x}=4\;\int_{}^{}{{\left|{{\zeta
}_{\;\overline{z}}}\;\right|}^{2}+{\left|{{\zeta
}_{\;z}}\;\right|}^{2} \over {\left({1+{\left|{\zeta
}\;\right|}^{2}}\right)}^{2}}\;d^{2}{x}, \eqno(5)$$ which
compared with Eq. (4) leads to the Bogomol'nyi inequality
$W \geq 4 \pi \;   {\left|\;Q\;\right|}$. When the equality
holds,  the field  $\zeta$ has  to be an
(anti)holomorphic function of $z$, leading to the
Belavin-Polyakov instantons [11].\par In our approach
 the spin $\vec S$ is represented by  the matrix $S =
S_{\alpha }\sigma _{\alpha } \;(\alpha  = 1,2,3)$, where
$\sigma _{\alpha }$ are the Pauli matrices.
Then, we
 diagonalize $S$  by  resorting to a right-invariant local
$U(1)$ transformation $g$,  such that $ S = g \sigma _{3}
g^{-1}.$ \noindent The matrix $g$ belongs   to  $U(2)$
 and has the form $$g=\pm\;
{\left[{2\left({1+{S}_{3}}\right)}\right]}^{-1/2}\left({\matrix{1+{S}_{3}&{-
{\overline{S}}_{+}}\cr
{S}_{+}&1+{S}_{3}\cr}}\right)exp\left({i{\sigma }_{3}{\eta
 \over 2}}\right),  \eqno(6)$$  where  $\eta$ is an
arbitrary real function. The  corresponding chiral current
$J_{\mu}\; (\mu = 0, 1, 2)$  is  decomposable into a
 diagonal and  an off-diagonal part:  $$ J_{\mu } =
g^{-1}\partial _{\mu }g = \left({{i}\over{4}}\right)\sigma
_{3}V_{\mu }  + \left({\matrix{0&-\;{\overline{q}}_{\mu }\cr {q}_{\mu
}&0\cr}}\right) .    \eqno(7)$$     By definition,   the chiral
current $J_{\mu}$ satifies  the zero curvature condition
$${\partial }_{\mu }{J}_{\nu }-{\partial }_{\nu }{J}_{\mu
}+\left[{{J}_{\mu },{J}_{\nu }}\right]=0 .
\eqno(8)$$  Its  off-diagonal and diagonal parts are
$$\eqalignno{&{D}_{\mu }{q}_{\nu }={D}_{\nu }{q}_{\mu },
& (9)\cr
&\left[{{D}_{\mu },{D}_{\nu
}}\right]=-{i \over 2}\left({{\partial }_{\mu }{V}_{\nu
}-{\partial }_{\nu }{V}_{\mu
}}\right)=-2\left({{\overline{q}}_{\mu }{q}_{\nu }-{q}_{\mu
}{\overline{q}}_{\nu }}\right), &(10)\cr} $$   where we
have introduced the  covariant derivative ${D}_{\mu
}={\partial }_{\mu }-{i \over 2}{V}_{\mu }$. \par
In the variables $q_{\mu}$, the equation of motion (1)
becomes      $$ q_{0}= i D_{m} q_{m},  \eqno(11)$$ which
allows us to eliminate the function $q_{0}$ from  Eqs.
(10-11).   Now, let us  introduce the  new fields    $$ \psi
_{\pm } = {(q_{1}\pm {\mit i}\;q_{2})\over{2 \sqrt {\pi}}},
\;\;  A_{0}= V_{0} - 8 (\mid \psi _{+}\mid ^{2}+
\mid \psi _{-}\mid ^{2}) ,\; \; A_{i} =  V_{i}\;
\;\;(i=1,2)
\eqno(12)$$    and,  correspondingly,  the new
covariant derivatives $D_{\mu } = \partial _{\mu } -
\left(i/2\right)  A_{\mu }$.
Then we notice that the
expressions of
 $\psi_{\pm}$ and $A_{j}$ in terms of
 $\zeta$  are   $$ \psi _{\pm } =
{1\over{2 \sqrt {\pi}}} \; {{\left( \partial _{1 } \,\pm\, i
\partial_{2}\right)\zeta  }\over{(1 +
 |\zeta |^{2})}} \; e^{ i\eta }, \qquad A_{j} = 2 i  \,
{{(\zeta \partial _{j}\bar{\zeta } - \bar{\zeta }\partial
_{j}\zeta )}\over{(1 +  |\zeta |^{2})}} + 2\partial _{j}\eta
.\eqno(13)  $$\par   For
$\mu=1$ and  $\nu=2$,   we see  that Eq. (9)
  takes the
form $$\gamma \equiv \left(D_{1} + i D_{2}\right) \psi_{-}
-  \left(D_{1} - i D_{2}\right) \psi_{+} = 0 .\eqno(14) $$
This equation can be  interpreted as a geometrical
constraint for the CS gauged   NLS  system,
 arising from the remaining   equations (9)  and
 Eq. (10), i.e.  $$\eqalignno{ &i{D}_{0}
{\psi}_{\pm}+\left({{{D}_{1}}^{2}+{{D}_{2}}^{2}}\right){\psi
}_{\pm }+8 \pi \, {\left|{{\psi }_{\pm }}\;\right|}^{2}\,{\psi
}_{\pm }=0 ,& (15.a) \cr &{\partial }_{2}{A}_{1}-{\partial
}_{1}{A}_{2} = 8
\pi
\left(\;J_{0}^{+} - J_{0}^{-}\right) ,& (15.b) \cr &{\partial
}_{0}{A}_{j}-{\partial }_{j}{A}_{0}= - 8 \pi \,{\varepsilon
}_{j\ell}\, {\left(\, J^{+}_{l} - J^{-}_{l} \right)}, &(15.c) \cr}$$
where  $$ J_{0} ^{\pm } =
  |\psi _{\pm }|^{2}, \qquad  J^{\,\pm }_{l} =   i\, (\psi
_{\,\pm } \bar{D}_{l}\bar{\psi }_{\, \pm } -
\bar{\psi }_{\,\pm } D_{l}\psi _{\,\pm })  \; \; \left( l = 1,
2\right). \eqno(16)$$
   Here we  introduce  the
charge density currents  satisfying
    the continuity equation
$\partial_{\mu}\,J^{\,\pm }_{\mu} = 0$. Then,   the
matter fields
$\psi_{\pm}$ carry a conserved electric charge and
we can interprete  the  quantity  $${\cal Q}  =  \int \left(
J_{0}^{+} -
J_{0}^{-}\right)\;d^{2}x  \eqno(17)$$  as the  (conserved)
  total electric charge.  In terms of
$\psi_{\pm}$ it is easy to see that
$Q$ and  $\cal Q$ coincide, thus  we are led to the
quantization of the total electric charge.
 Furthermore,  Eq.   (15.b)  implies
 the  quantization of the magnetic flux: $\Phi = \int B
d^{\,2}x =  \int \left({\partial }_{1}{A}_{2}-{\partial
}_{2}{A}_{1}\right) d^{\,2}x = - 8  \pi {
Q}$. \par
 Moreover,  the continuity equations for the currents
$J^{\pm}_{\mu}$ predict that   the  total number of
particles (anyons)
$${\cal N}  =    \int \left( J_{0}^{+} + J_{0}^{-}\right)d^{2}x
\eqno(18)$$ is  also a conserved quantity.  Now,
comparing  Eq. (5) with (18) , we obtain  $ {W} = 4 \pi
  {\cal N}$.  Then,  in the tangent space formulation,   the
Bogomol'nyi  inequality takes the form $\left| Q\;\right|
\leq {\cal N}$.  Of course,  the equality holds    when
$\psi_{+}$, or $\psi_{-}$,   vanishes.    In such a  case,  the
system (14 - 15)  reduces to the self-dual CS model [2].
 This model is completely
determined in terms of the solutions of  the Liouville
equation.  This simple observation enables us
to identify the magnetic vortices, or equivalently the
instantons of Belavin-Polyakov type,  with the solitons of
the self-dual CS model, that is  with  the anyons. In
particular, from (17) we have an explicit correspondence
between the magnetic topological charge and  the anyon
electric charge,  whose quantization finds here a
geometrical interpretation. In ref. [8]   the  more general
static case ${\left( D_{1} + i D_{2} \right)}
\; \psi _{-} =   {\left( D_{1} - i D_{2} \right)} \; \psi _{+} =
0$  is treated.   This  leads to a  special reduction of
the euclidean conformal affine Toda system, which
contains the sinh-Gordon, or the Poisson-Boltzmann
equation, in a  particular gauge.  \par  Finally,
we notice that   the constraint  (14) is identically satisfied
  using the expressions  (13). Indeed, we find  the
identity  $$ {\left( D_{1} + i D_{2} \right)}
\;
\psi _{-} =   {\left( D_{1} - i D_{2} \right)} \; \psi _{+} = {1
\over 2}\; \left[ \; {\zeta }_{\;z\;\overline{z}}  -
2{{\zeta }_{\;z}\;{\zeta }_{\;\overline{z}} \over
1+{\left|{\zeta }\;\right|}^{2}}\overline{\zeta }\; \right]\;
{e^{i\eta } \over {1 +  \left|\zeta \;\right|^{2}}}.
\eqno(19) $$  Equating to  zero  the quantity in the square
brackets, we  recognize   the stationary LLE    in the
stereographic variable (see Eq. (3)). Thus, all static
solutions to Eq. (1) are mapped, modulo gauge
transformations, into solutions to  the sinh-Gordon
reduction  studied in       ref.   [8].
\par\medskip
\noindent {\bf 3.} $\qquad$
System (15) can be derived from the Lagrangian
 $$ {\cal L} ={i\over  2}\; (\bar{\psi } D_{0}\psi  -
\bar{D}_{0}\bar{\psi } \psi  ) - \bar{D}_{a}\bar{\psi }
D_{a}\psi  + 4 \pi \;(\bar{\psi } \sigma _{3}\psi
)(\bar{\psi } \psi ) + {1\over 32 \pi} \; \epsilon ^{\mu \nu
\lambda }A_{\mu }\partial _{\nu }A_{\lambda }\;\; ,
\eqno(20) $$  \noindent where  we have used the spinor
notation  $\psi ^{T} = (\psi _{+},  \psi _{-})$, with
$\bar{\psi } = \psi ^{+}\sigma _{3}$,   and the summation
over $a = 1, 2$ is understood. Let
us observe  that   the strength of
the self-interaction  and   the CS coupling are fixed by the
original spin model. The specific ratio between these
constants allows  us  to  perform the self-dual
reduction, which  is  a very  special subcase in  more general and the
phenomenological theories of CS type   [2].
Moreover,  in the effective theories for  FQHE  the
CS coupling is related to the so-called
{\it filling factor} $\nu$, which in  the present case is  $
1/2$, where  the units $ \hbar = c = 1$ and
$e = 1/2$ have been used. This result   is  consistent with
the so-called chiral spin fluid  model [1 - 12]. Moreover, recent
experimental observations in double-layer samples
indicate the compatibility  of
$\nu$, with   values  which are inverse of even numbers  [13].  \par A
final remark     concerns the parity  invariance of the
Lagrangian (20),
 which  changes the total sign under  the  transformation
$$\matrix{{x}_{1}\rightarrow x{}_{1},&{x}_{2}\rightarrow
-x{}_{2}\cr {\psi }_{+}\leftrightarrow {\psi
}_{-},& D_{1} + i D_{2}\leftrightarrow D_{1} - i D_{2}\;.\cr}
\eqno(21)$$ So that  equations of motion and  the constraint (14)
 are both  left
invariant. This fact is an
obvious consequence of the parity invariance of the spin
model (1) and it is another example,  against the common
belief, that the presence of the CS interaction    breaks  the
P and T invariance. \par Now let us enter into the detail of
the hamiltonian structure of  the model \hbox{(14 - 15)}
by resorting to the
 first order Lagrangian method  proposed in ref. [14],
which implements  the classical  Dirac   method for
constrained Hamiltonian systems [15].\par  First, we
consider as dynamical variables  $\xi  = \left(\psi _{+} ,
\bar{\psi }_{+}, \psi _{-} , \bar{\psi }_{-} , A_{1} ,
A_{2}\right) $  and, defining  $  a = ({i\over{2}}\,\bar{\psi
}_{+} ,-{i\over{2}}\,\psi _{+} ,-{i\over{2}}\,\bar{\psi }_{-} ,
{i\over{2}}\,\psi _{-} , {1\over{32 \,\pi}}\, A_{2} ,- { 1\over{32\,
\pi}}
\,A_{1})$,   we rearrange  (20)  in the form
\par$$\eqalignno{& {\cal L'} = a_{m} \dot{\xi} ^{m} - {\cal
H}(\xi )   + {1\over{2}}
 A_{0} \Gamma_{1} , &(22)$$ \cr   &
{\cal H} = \bar{D}_{a}\bar{\psi } D_{a}\psi  - 4
\pi \;(\bar{\psi }
\sigma _{3}\psi )(\bar{\psi } \psi ) &(23)\cr &  \Gamma
_{1}  \equiv   {{\delta {\cal L}}\over{\delta A_{0}}} =
\bar{\psi } \psi   + {1 \over{8\, \pi}}\;
\epsilon ^{ij} \partial _{i} A_{j}   =   0.  &(24)\cr}$$
Thus   $\cal L'$   describes the system (15 a,c)
constrained  by Eq. (24), that is the
Chern-Simons version of  the Gauss law  (15.b). In
(22)  $A_{0}$ plays the role of Lagrange multiplier.
Furthermore, $\cal H$  would be the density of the
Hamiltonian $H$.  In fact,   the  corresponding
Euler-Lagrange equations take the hamiltonian  form
$$\dot{\xi}^{i} = {f_{i j}}^{-1}{{ \partial
{\cal H}} \over{ \partial \xi^{j}}}  = \{  \xi^{i}, { H}\,\},
\eqno(25)$$
 where  $     f_{ml}=\partial
a_{l}/\partial \xi ^{m} - \partial a_{m}/\partial \xi ^{l}
$   is   a nonsingular symplectic matrix and we
have defined  the Poisson brackets   by
$
\{\xi ^{m},\xi ^{l}\} = f^{-1}_{m l} \delta^{2}\left({\bf x} -
{\bf y}\right)$. Specifically, in our case  the nonvanishing
Poisson brackets  are  $$ \{\psi _{\pm }({\bf x}),\bar{\psi
}_{\pm }({\bf y})\} = \mp \,i\;\delta ^{2}({\bf x} - {\bf
y}),\quad \{A_{i}({\bf x}),A_{j}({\bf y})\} = 16 \pi\,
\epsilon _{ij}\delta ^{2}({\bf x} - {\bf y}) .\eqno(26)$$   By
using  Eq. (33), one easily shows  that $ \dot{\Gamma
}_{1} =
\{\Gamma _{1},{H}\} =  0 $;  consequently,  in  phase
space the Gauss'  law defines an  invariant  submanifold
under the   dynamics. On the other hand,   Eq.  (14)
provides the   further complex primary constraint  $
\gamma  = 0$,  whose  real and immaginary parts will be
denoted by
$\Gamma _{2}$ and $ \Gamma _{3}$.   By evaluating   the
Poisson brackets,  we find  that
 $ \{\gamma ,  H \} \approx  0
$, i.e. $\dot{\gamma}$ is vanishing only on the
submanifold   defined by (14)  or,  in other words,
$\Gamma_{1}$ and  $\Gamma_{2}$  are weakly invariant
under time evolution.  Furthermore,  they do not produce
secondary constraints.  However,  since
  $\{\Gamma_{1},  \Gamma_{2}, \Gamma_{3}\}$ is a set  of
first class  constraints (i.e., $\{\Gamma_{i},{ H}\}
\approx 0$ and
 ${\rm det} \left( \{\Gamma_{i},
 \Gamma_{j}\}\right) \approx 0$ ),   we need a
gauge-fixing condition,  for example the   Coulomb  gauge
condition    $ \Gamma _{4} = \partial _{i}A^{i}\approx  0,
$  which is  the most  used in the litterature [16,
17,  18].  So,  we are able  to compute the  Dirac
brackets,  defined  by  $ \{\xi ^{m},\xi ^{l}\}_{D} \equiv
\{\xi ^{m},\xi ^{l}\} - \{\xi ^{m},\Gamma _{s}\}({d^{-
1}})_{ss'}\{\Gamma _{s'},\xi ^{l}\} $,  where  $ d_{ s s' }   =
\{\Gamma _{s},\Gamma _{s'}\} $.  In our case,  the
non-vanishing    matrix  elements $d_{s s'}$ are
distributions in terms of  $\delta ^{2}({\bf
x}-{\bf y})$  and its derivatives. Then,  inversion    can be
performed
  only in implicit form.   In order to    avoid  this  type of
problem,   one  can  look  for    a  different  gauge-fixing
condition.  For instance, a more  general non-local gauge
fixing can be choosen   [19], or     the so-called
"axial gauge" [20].   Nevertheless, in all generality
   the corresponding Dirac brackets  are    rational  in
the  field  variables.  Then    a    quantization of the
model  seems to be rather difficult, because of the
problem arising in the ordering and the  inversion of the  operators
involved.  This  is one of the main  reasons which
address  us to look for  an alternative  approach, which is
invariant  under  the choice of the gauge. \par \noindent
 ${\bf 4.} \qquad$  In this Section, we will adopt a
slightly different  point of view, in order to tackle  the
above mentioned  problems.
Moreover,   we would like  to  understand the quantum
meaning of the geometrical constraints
$\gamma  $,  as seen  separately from the basic
CS-NLS  field  theory.  Then, a possible route
consists in  quantizing   system (14 - 15) in two steps: \par
\noindent 1. first,      we  apply  the  procedure  of  the
gauge-invariant quantization to system (15)  following
the approach suggested in  ref. [21]; \par \noindent 2.
then,  we use  the constraint
$\gamma $    to  select the physical states among  all the
gauge invariant quantum states. \par First of all,  let us
change    the hamiltonian description  by   taking   $${\cal
H'} = {\cal H}\left(\xi\right) + f_{0} \Gamma_{0} + {1\over
{2}} A_{0}
\Gamma_{1},\eqno  (27)$$  where we have as primary
constraint $\Gamma_{0} \ \equiv  \pi_{0} = 0$,  $\pi_{0}$
being the  momentum conjugated to the canonical variable
$A_{0}$, such that $\{ A_{0}, \pi_{0}\} =
\delta^{2}\left({\bf x} - {\bf y} \right)$.  This  constraint
is implicit in the previous formulation, since $\partial {\cal
L}/ \partial \dot{A_{0} } = 0$.
  $\Gamma_{1}$ is  now a secondary constraint, arising
from the Poisson bracket $\{ \pi_{0}, {H}\}$. Finally,  in
the expression (27)  $f_{0}$ is an  arbitrary  function,
which defines  the evolution of $A_{0}$.  Furthermore,
noticing that  the special    canonical pair
$\left( A_{1}, A_{2}\right )$  breaks the rotational
covariance of the theory, it is more  convenient   to
consider
 the  decomposition  of the  vector potential into   a
longitudinal and   a   transverse part
$A_{i}({\bf x}) = A^{L}_{i} + A^{T}_{i} =  2\partial _{i}\,\eta
({\bf x}) + \epsilon _{ij}\partial _{j}\chi ({\bf x})
$.  Recalling that the
magnetic field  is  $ B = \epsilon ^{ij}\partial _{i}A_{j} = -
\partial ^{2}\chi  $,    we derive  formally the useful
relation
$$ A^{T}_{i}({\bf x}) =  - \epsilon _{ij}(\partial
^{-1}_{j}B)({\bf x}), \eqno(28)$$ where we have defined
  $\partial^{-1}_{j}f \left({\bf x}\right)
=  {1\over{2\pi}}
\partial^{\left(x\right)}_{j}\int
  \ln |{\bf x}-{\bf y}| f({\bf y}) d^{2}y. $
Analogously, it is convenient to express $ \psi _{\pm }$ in
terms of the
 canonical variables  $\left(P_{\pm } ,  Q_{\pm }\right)$
by   $$ \psi _{\pm } = {1\over{\sqrt{2}} }\left(P_{\pm }
\mp  i Q_{\pm }\right).\eqno(29)$$  Hence,  the Poisson
structure associated with  the hamiltonian system  given
by  (27)  is
    $$ \matrix{\{Q_{\pm }({\bf
x}),P_{\pm }({\bf y})\} = \delta ^{2}({\bf x} - {\bf y}),
\;\{A_{0}({\bf x}),\pi _{0}({\bf y})\} = \delta ^{2}({\bf x} -
{\bf y}), \cr \{B({\bf y}),\eta ({\bf x}) \} =
8\,\pi\;\delta ^{2}({\bf x} - {\bf y}).\cr}\eqno(30) $$
\noindent Furthermore, the equations  of motion are
supplemented by   the  first class constraints
$\Gamma_{0}  \approx
\Gamma_{1} \approx 0$.
\par   Now,  we are look for  a suitable canonical
transformation, such that some of the new momenta are
equal to  the constraints. In such a way, the  coordinates
canonically conjugated to these momenta  have an
arbitrary  time evolution and  are remnants of  the gauge
invariance of the theory.  The remaining canonical
variables  define the  prominent dynamical degrees of
freedom of the theory.  \par  First,  let  us denote by  $
(Q^{*}_{\pm },P^{*}_{\pm }), (A^{*}_{0},\pi ^{*}_{0}), (\eta
^{*},\pi ^{*}_{1}) $    new  canonically conjugated
coordinates, where  the momenta
$\pi^{*}_{i}$  have to be set  equal to the constraints
$\Gamma_{i}$.   This can be done through  a suitable
generating function
$F =F\left(Q_{\pm },A_{0},\eta  ; P^{*}_{\pm },\pi
^{*}_{0},\pi ^{*}_{1}\right)$, satisfying the set of
generalized Hamilton-Jacobi equations $$\pi^{*}_{i} =
\Gamma_{i}\left( Q_{\pm },A_{0},\eta ; {\partial
F\over{\partial Q_{\pm }}},  {\partial F\over{\partial
A_{0}}}, {\partial F\over{\partial \eta}}\right).
\eqno(31)$$
 The integrability of this  system is assured by $\{
\Gamma_{0}, \Gamma_{1}\} = 0$, moreover    its general
solution  is given by      $$\eqalign{F={\pi }_{0}^{{}^*}{A}_{0}+&\left({{\pi
}_{1}^{{}^*}+{1
\over 2}{P}_{+}^{{}^*2}-{1 \over 2}{P}_{-}^{{}^*2}}\right)\eta\; + \cr
&\int\sqrt {{P}_{+}^{{}^*2}-{Q}_{+}^{2}}d{Q}_{+}+\int{\sqrt
{{P}_{-}^{{}^*2}-{Q}^{2}}d{Q}_{-}} .\cr} \eqno(32)$$  The
corresponding   canonical transformation  reads  $$
\cases{\pi _{0} = \pi ^{*}_{0} ,\; A_{0} = A^{*}_{0} ,\quad
-{1 \over{8\, \pi}} B =
\pi ^{*}_{1}+ {1\over{2}} P^{*2}_{+} - {1\over{2}} P^{*2}_{-}
,\;
\eta  = \eta ^{*}, \cr
\cr  Q_{\pm } = P^{*}_{\pm } \sin \left(\,{Q^{*}_{\pm }
\over{P^{*}_{\pm }}} \mp  \eta ^{*}\right) , P_{\pm } =
P^{*}_{\pm } \cos \left(\,{Q^{*}_{\pm }\over{P^{*}_{\pm }
}}\mp  \eta ^{*}\right).}\eqno(33) $$ \noindent Finally,
the  new gauge-invariant degrees of freedom are
$$
\Phi _{\pm } = {1\over{\sqrt{2}}} P^{*}_{\pm } \exp
\left(\mp\,  i \,{ Q^{*}_{\pm }\over{P^{*}_{\pm }}}\right) =
\psi_{\pm} \exp\left( - i \eta\right) , \qquad
\pi^{*}_{\pm} =\overline{\Phi}_{\pm}, \eqno(34)$$
\noindent which fulfill the   canonical brackets  \par $$
\{\Phi _{\pm }({\bf x}),\overline{\Phi }_{\pm }({\bf y})\}
= \mp \, i \, \delta ^{2}({\bf x} - {\bf y}) . \eqno(35)$$
\noindent In terms of these new variables,  the
Hamiltonian density   becomes
$$ \matrix{ {\cal H} = (\partial _{a}+ {i\over{2}}
A^{T}_{a})\overline{\Phi }_{+}(\partial _{a}- {i\over{2}}
A^{T}_{a})\Phi _{+} - (\partial _{a}+ {i\over{2}}
A^{T}_{a})\overline{\Phi }_{-}(\partial _{a}- {i\over{2} }
A^{T}_{a})\Phi _{-}  \cr \cr  - 4\,\pi\; (|\Phi _{+}|^{4}-
 |\Phi _{-}|^{4})] + f^{0} \pi^{*}_{0} +
{1\over{2}}A_{0}^{*}\pi^{*}_{1},\cr}\eqno(36) $$  where
\hbox{$A^{T}_{a}({\bf x}) = 8\, \pi \; \epsilon _{a b
}\partial ^{-1}_{b}(|\Phi _{+}|^{2}- |\Phi _{-}|^{2})({\bf x}). $}
Furthermore, the constraint $\gamma$ takes the form$$
\gamma  = (\partial _{+}- {i \over{2}} A^{T}_{+})\Phi _{-} -
(\partial _{-}- {i \over{2}} A^{T}_{-})\Phi _{+},
\eqno(37)$$ where $ \partial_{\pm} =
\partial_{1} \pm i\,\partial_{2} $ and  $ A_{\pm}^{T} =
A_{1}^{T} \pm i\,  A_{2}^{T}$. \par
 Now, let us observe that  the relations (13) are
  algebraically invertible in terms of gauge invariant
quantities when the topological density is non vanishing,
namely $$ \zeta  = {i \over{4\, \sqrt{\pi}}}{{\Phi
_{-}\;A^{T}_{+} - \Phi
_{+}\; A^{T}_{-}}\over{|\Phi _{+}|^{2}- |\Phi
_{-}|^{2}}}.\eqno(38) $$  Then,  we can
directly reconstruct the spin field in terms
of $\Phi_{\pm}$, although it turns out to be  a  non local
field  in view  of the expressions  for $A^{T}_{\pm}$. In this
regard,  let us  observe    that,   in the one-dimensional
completely integrable case,  the   relation  between
$\zeta$ and   the NLS  matter field is also analytic in
nature,  but we need to  solve the associated linear
spectral problem. \par  Furthermore,   keeping in
mind that the current densities (16) are manifestly gauge
invariant, we can give the expressions of the  linear and
angular momentum for the model  (14 - 15) in  a  gauge
invariant form. \par Firstly,   in opposition to  what
happens for  the electric charge densities, the   linear
momenta  $ P^{\pm }_{m} =\int J^{\pm }_{m}\, d^{2}x $,
associated with  the
$\Phi_{\pm}$ fields separately,  are  no   longer conserved,  but they
evolve according to   Newton's law  $${\rm  {d\over {d t}}} P^{\pm
}_{m} =
   8 \; \epsilon _{m n }\int (J_{0} ^{-}J^{+}_{n}   -  J_{0}
^{+}J^{-}_{n}) d^{2}x. \eqno(39) $$ \noindent We notice
that     both  $\Phi_{\pm} $   are subjected  to
the same force. Therefore,  the  total kinetic
momentum
$${P}_{m}={P}_{m}^{+}-{P}_{m}^{-}=\int\left(\;{{\overline{\Phi
}}_{+}\partial_{m} {\Phi }_{+}-{\overline{\Phi
}}_{-}\partial_{m} {\Phi }_{-}}\right){d}^{2}x+c.c.
\eqno(40)$$ is conserved, the $"-"$ comes  from the
"negative mass" associated to $\Phi_{-}$          .  The total
angular momentum  $ M = M_{+}- M_{-}$ is also conserved
,  with
$${M}_{\pm }=\int \left({{x}_{1}{J}_{2}^{\pm} -
{x}_{2}{J}_{1}^{\pm }}\right){{d}^{2}x}= {\ell }_{\pm
}+{s}_{\pm }
   \eqno (41)$$  and $$\matrix{l_{\pm} =& - i \; \int {{\bar
\Phi}_{\pm}\left(  x_{1} \partial_{2} -  x_{2}
\partial_{1}\right) \Phi_{\pm} } d^{2}x + c.c. ,\cr
 s_{\pm} =&  \;
\int { |\Phi_{\pm}|^{2}\left(   A^{T}_{1}\, x_{2} -
 A^{T}_{2} \,x_{1}  \right)  } d^{2}x . \cr} \eqno(42)$$
The quantities $ l = l_{+} - l_{-}$ and $ s = s_{+} - s_{-}$
are interpreted as  orbital and spin angular momentum,
respectively.  In particular,  the general result  $s =  2  \,
Q^{2}$  suggests that  vortices have non zero momentum
also in the static case.\par
 Finally, since  the magnetization $\vec {\cal M }= \int
{\vec S}\; d^{2}x$  is a conserved non-local quantity for
the  LLE system, via Eqs. (38) and (2) we find a further
 gauge invariant conserved quantity for the system (14 -
15) ,  related to  the  original global
$SU(2)$  symmetry.
\par  We are now  ready to pass on the
quantum theory in a straightforward manner, just  by
replacing  the canonical brackets  by  $i$ times the
commutators (or the  anti-commutators, if the matter
fields are supposed to be fermionic).  At    the quantum
level, the first class constraints $\Gamma_{i}$  become
the operators  ${\hat{\Gamma }}_{i}$,  which  must
annihilate  the physical states. This fact implies that the
quantum states are independent  of  $A_{0}$ and
invariant under time-independent
gauge-transformations.   Therefore, the gauge - invariant
operators  of the theory must commute with
${\hat{\Gamma }}_{i}$.  Then,  recalling the Poisson
brackets for the Hamiltonian system described by  (36),
we can  establish the correspondence
${\Phi}_{\pm }\rightarrow {\hat{\Phi}}_{\pm }$ and
${\overline{\Phi}}_{\pm }\rightarrow
{\hat{\Phi}}^{\dagger}_{\pm }$, with the equal time (anti-)
commutation relations    \par $$ [{\hat \Phi} _{\pm}({\bf
x}),{\hat{\Phi }}^{\dagger}_{\pm}({\bf y})]_{\pm }= \pm\,
\delta ^{2}({\bf x} - {\bf y}). \eqno(43) $$  In the physical
gauge-invariant subspace of the full Hilbert space, the
Hamiltonian  is given by  $$\matrix{ { \hat{H}}
=\int\{-{\hat \Phi} ^{\dagger}_{+}(\partial _{a}- {i
\over{2}} A^{T}_{a})^{2}{\hat \Phi} _{+}  +  {\hat \Phi}
^{\dagger }_{-}(\partial _{a}- {i \over{2}}
A^{T}_{a})^{2}{\hat \Phi} _{-}
 \cr \cr  -  4\, \pi\; :({\hat \Phi} ^{\dagger}_{+}{\hat \Phi}
_{+})^{2}- ({\hat \Phi} ^{\dagger}_{-}{\hat \Phi}
_{-})^{2}:\}d^{2}x , \cr}\eqno(44)$$ where  the  normal
product  operator $ : \;\;:$   and  $A^{T}_{a} =  8\,
\pi \,\epsilon _{a b }\,\partial ^{-1}_{b}(  \hat \Phi
_{+}^{\dagger}  \hat
\Phi_{+} - \hat \Phi _{-}^{\dagger}
\hat  \Phi_{-})({\bf x}) $  have been used.  The fields
${\hat \Phi}_{\pm}$ are  local and invariant under local
gauge transformations generated by the first-class
constraints. Therefore,  they  may  be exploited to study
the breaking of global simmetries. \par
 The quantum version of relations (30) leads
to
  \hbox{$ [{\hat B}({\bf x}),{\hat \eta} ({\bf y})] =
8 \pi i\delta ^{2}({\bf x} - {\bf y})$},  which implies
the relation
  $$ \exp \left(i\,{\hat \eta}  \left({\bf
y}\right)\right){\hat B}\left({\bf x}\right)\exp
\left(-i\,{\hat \eta}  \left({\bf y}\right)\right) = {\hat
B}\left({\bf x}\right) + 8\, \pi \,\delta ^{2}\left({\bf x}
- {\bf y}\right).\eqno(45) $$\noindent  This result is used
to prove    that $ {\hat \Phi}_{\pm} $  are  the
gauge  invariant   operators   which    create  a
charge-solenoid composite,  having magnetic  flux  equal
to $\pm  8 \, \pi \, $.
\par  The Hamiltonian
 (44) commutes with the  particle
  number operators \par $$ {\hat N}_{\pm }= \int \, {\hat
\Phi} ^{\dagger}_{\pm }{\hat \Phi} _{\pm }d ^{2} x
.\eqno(46) $$  Hence, we can
 diagonalize them  simultaneously via  the  Fock
representation for  a quantum state with    $N = N_{+} +
N_{-}$  particles, which is  given  by
$$ \eqalign{|N_{+},
N_{-}>&  = \int_{}^{}\prod\nolimits\limits_{i=1}^{{N}_{+}}
{d}^{2}{x}_{i}^{+}\prod\nolimits\limits_{j=1}^{{N}_{-}}
{d}^{2}{x}_{j}^{-}\Psi \left({\bf
x}^{+}_{1},\dots  , {\bf x}^{+}_{N_{+}},{\bf x}^{-}_{1},\dots  ,
{\bf x}^{-}_{N_{-}}\right) \cr &{\hat \Phi}
^{\dagger}_{+}\left({\bf x}^{+}_{1}\right)\ldots  {\hat \Phi}
^{\dagger}_{+}\left({\bf x}^{+}_{N_{+}}\right) {\hat \Phi}
^{\dagger}_{-}\left({\bf x}^{-}_{1}\right)\ldots  {\hat \Phi}
^{\dagger}_{-}\left({\bf x}^{-}_{N_{-}}\right) |{\bf 0}>,
\cr}\eqno(47)$$   where the vacuum state  is defined by
${\hat
\Phi} _{\pm }|{\bf 0}> = 0$.   The function  $\Psi $ is an
arbitrary element of  the Hilbert space  ${\sl L}_{2} [{\cal
R}^{2\left(N_{+} + N_{-}\right)}]$  obeying   the
Schr\"odinger equation for   $\left(N_{+}
+N_{-}\right)$-bodies.
  Furthermore,  the state (47) is gauge
invariant by construction.
Thus, we have
  selected in the full Hilbert space  only those  states
which  contain the whole  physical information. \par
Nevertheless, up to now we have dealt with a quite general
"anyonic" field theory, without using  the constraint
$\gamma$, which in some sense  takes into account the
properties of the original spin field model. As a
consequence, "quantizing" Eq. (37) in the form $$\hat
\gamma = \hat D_{+} \hat \Phi_{-}\left({\bf x}\right) -
\hat  D_{-} \hat \Phi_{+}\left({\bf x}\right), \eqno(48)$$
we should obtain the physical  states by  requiring that
$\hat \gamma |N_{+}, N_{-}> = 0 $. However,  for arbitrary
occupation numbers  $\left( N_{+}, N_{-}\right)$ this
equation leads  to complicated relations for the wave
function
$\Psi$. Here we display the result only for the finite
boson  case
$\left|{{N}_{+},0}\right\rangle$, namely
$${\Psi }\left({{\bf x}_{\rm 1}^{+},\cdots,{\bf x}_{\rm
{N_{+}}}^{+}}\right)={\cal F}
\left({{\overline{z}}_{1}^{\;+},\cdots,{\overline{z}}_
{{N}_{+}}^{\;+}}\right)\prod\nolimits\limits_{i\,<\,j} {\left|{{\bf
x}_{\rm i}^{+}-{\bf x}_{\rm j}^{+}}\right|}^{- \,2 },
\eqno(49)$$ where $\cal F$ is a  holomorphic function of
its arguments. By using a singular gauge transformation
[23-24],  the expression (49)   takes the  form
 of the Laughlin  wave function  [25] , describing  a condensate state of
bosonic solitons, which may be related with quantum disorderd state of
the original ferromagnets [12]. Furthermore, it should be proved that the
quantum state corresponding to the wave function (49) is actually an
eigenvector of the Hamiltonian (44). This work is in progress, and we
have strong indications in the positive sense, because at the classical
level the Hamiltonian (23) can be written as a linear combination of the
constraint
$\gamma$ and
$\Gamma_{1}$. \par This work was supported in part by MURST
of Italy and by INFN - Sezione di Lecce. One of the authors (O. K. P.) thanks
the Department of Physics of Lecce University for the warm hospitality. \vfill
\eject \par  {\bf References}\par \medskip\medskip \noindent \item{[1]} F.
 A. Zee,  {\it Prog. Theor. Phys. Suppl.
}{\bf107}, 77 (1992).\par \noindent  \item{[2]}  R. E. Prange and S.
M. Girvin, {\it  The Quantum Hall Effect}  (Springer-Verlag,
New York, 1990). \par \noindent  \item{[3] } S. C. Zhang, T. H.
Hansson and S. Kivelson, {\it Phys. Rev. Lett.} {\bf 62}, 82
(1989). \par \noindent  \item{ [4]}
 Z. F. Ezawa, M. Hotta and
A. Iwazaki,{\it Progr. Theor. Phys. Suppl.}  {\bf 107}, 185
(1992).\par
\noindent  \item{[5]}  R. Jackiw and S-Y. Pi, {\it Phys. Rev. Lett.}
{\bf 64}, 2969 (1990); {\bf 66}, 2682 (1990);  {\it Phys.
Rev.} {\bf D 42}, 3500  (1990).   \par
\noindent  \item{[6] }  L. Martina, O.K. Pashaev, G. Soliani, {\it
Mod.  Phys. Lett.}  {\bf A 8}, 3241 (1993). \par \noindent
 \item{[7]}  V. G. Makhankov and O.K. Pashaev, {\it Integrable
Pseudospin Models in Condensed Matter}, in Sov. Sci. Rev.,
Math. Phys. {\bf 9}, 1  (1992).
 \par \noindent  \item{ [8]}  L. Martina, O.K. Pashaev, G. Soliani,
{\it Phys. Rev. } {\bf B  48},  15787 (1993). \par \noindent
 \item{[9]}  T. H. O'Dell, {\it Ferromagnetodynamics, the dynamics
of magnetic bubbles, domains and domain wall}, New York,
Wiley (1981). \par \noindent  \item{[10]}   N. Papanicolau and T.
N. Tomaras, Nucl. Phys {\bf B 360}, 425 (1991). \par
\noindent  \item{[11]} A. A. Belavin and A. M. Polyakov,{\it JETP
Lett.} {\bf 22}, 245 (1975). \par \noindent  \item{[12]} A. Zee,  {\it
Physics in (2+1)-dimensions} Y.M. Cho ed. (World Scientific, Singapore,
1992).
\par
\noindent
\item{[13]} M.  Greiter, X.G. Wen and F. Wilczek, Nucl. Phys. {\bf B374},
567 (1992),  "Paired Hall states in double layer electron
system", IASSNS-HEP-92/1 \par \noindent   \item{[14]}  L.
Faddeev, R. Jackiw, {\it Phys. Rev. Lett. } {\bf 60},692
(1988).\par \noindent  \item{[15]} P. A. M. Dirac, {\it Lectures on
Quantum Mechanics} (Yeshiva University, New York,
1964). \par \noindent
\par
\noindent  \item{[16]} P. de Sousa Gerbert, {\it Phys. Rev.} {\bf D
42}, 543 (1990).\par
\noindent  \item{ [17]}   T. Matsuyama, {\it Phys. Lett.} {\bf B
228}, 99 (1989). \par
\noindent  \item{[18]} C. G. Han, {\it Phys. Rev.} {\bf D 47}, 5521
(1993).\par \noindent  \item{[19]}  R. Banerjee, A. Chatterjee and
V. V. Sreedhar,  {\it Ann. Phys.}{\bf 222},254 (1993). \par
\noindent  \item{[20]}  M. Chaichian and N. F.  Nelipa, {\it
Introduction to gauge field theory} ( Springer-Verlag,
Berlin 1984).\par \noindent  \item{[21]} D. Boyanovsky, E. T.
Newman and C. Rovelli, {\it Phys. Rev. } {\bf D 45}, 1210
(1992);  D. Boyanovsky, {\it Int. J. Mod. Phys.} {\bf A 7},
5917, (1992).\par \noindent  \item{[22]} N. K. Pak  and R.
Percacci, {\it J. Math. Phys.} {\bf 30}, 2951 (1989); N. K.L
Pak, {\it Progr. Theor.  Phys.} {\bf 88}, 585 (1992).
\par \noindent  \item{[23]}  F. Wilczek, {\it Phys. Rev. Lett.} {\bf 48},
1144 (1982). \par \noindent  \item{[23]} S. Girvin et al., {\it Phys. Rev.
Lett.} {\bf 65}, 1489, (1992).
\par
\noindent
\item{[25]}  R.B.  Laughlin, {\it Phys. Rev. Lett.} {\bf 50}, 1395
(1983).\end